\documentclass[a4paper,12pt]{article}

\usepackage[left=2.5cm,right=2.5cm,top=2.5cm,bottom=2.5cm]{geometry}
\usepackage{graphicx}
\usepackage{amssymb}
\usepackage{multirow}
\usepackage{amsmath}
\usepackage{psfrag}
\usepackage{setspace}
\usepackage{subcaption}
\usepackage[colorlinks=true,bookmarks=true]{hyperref}
\usepackage{float}
\usepackage{bbm}

\hypersetup{citecolor=blue}

\newcommand{\dif}{\mathrm{d}}

\newcommand{\overbar}[1]{\mkern 1.5mu\overline{\mkern-1.5mu#1\mkern-1.5mu}\mkern 1.5mu}

\newcommand{\fracb}[2]{\displaystyle{\frac{#1}{#2}}}
\newcommand{\Eqref}[1]{(\ref{#1})}
\newcommand{\half}{\frac{1}{2}}

\newcommand{\brac}[1]{\left(#1 \right)}
\newcommand{\sbrac}[1]{\left[#1\right]}

\newcommand{\eq}{\,=\,}

\begin{document}

\title{Geodesic motion in the vacuum C-metric}
\author{Yen-Kheng Lim\footnote{E-mail: phylyk@nus.edu.sg}\\\textit{Department of Physics, National University of Singapore,}\\\textit{Singapore 117551}}

\date{\today}
\maketitle
 \begin{abstract}
 Geodesic equations of the vacuum C-metric are derived and solved for various cases. The solutions describe the motion of timelike or null particles with conserved energy and angular momentum. Polar, nearly-circular orbits around weakly accelerated black holes may be regarded as a perturbation of circular Schwarzschild geodesics. Results indicate that circular Schwarzschild geodesics of radius $r_0>6m$ are stable under small uniform accelerations along the orbital plane. These stable orbits undergo small oscillations around $r_0$, behaving like a harmonic oscillator driven by a periodic force plus another constant force. Circular orbits with axis parallel to the direction of black hole acceleration are also considered. In this case an algebraic relation expressing the condition of stability is obtained. This refines the stability analysis done in previous literature. We also present an analysis of radial geodesics along the poles. There exist a solution where a particle remains at unstable equilibrium at a 
fixed distance directly behind the accelerating black hole. Examples of numerical solutions are presented for other more general cases.
\end{abstract}

\section{Introduction} \label{intro}
The C-metric is a well known solution to Einstein's equation that describes black holes with mass parameter $m$ under uniform acceleration parametrized by $A$. The physical interpretation of the C-metric was obtained by, among others, Kinnersley and Walker \cite{PhysRevD.2.1359}, and Bonnor \cite{Bonnor:1983}. The uniform acceleration of the black hole is caused by conical singularities that may be interpreted either as the black hole being pulled by a cosmic string, or pushed by a cosmic strut.
\par
The most familiar form of the metric is perhaps written in C-metric coordinates (see Eq.~\Eqref{general_form} below), with structure functions $F(y)$ and $G(x)$ which are, respectively, polynomials in $y$ and $x$. In Ref.~\cite{Hong:2003gx}, Hong and Teo presented the C-metric in a new form in which $F(y)$ and $G(x)$ are factorized, allowing the roots of the functions to be written in simple forms. In our study of geodesic motion, it is often convenient to use spherical-type coordinates, where we will use the form presented in \cite{Griffiths:2006tk,Griffiths:2009dfa}. One of the advantages of using spherical-type coordinates is that the metric directly reduces to the Schwarzschild limit when $A\rightarrow 0$. This will allow direct comparisons between orbits of the C-metric and the Schwarzschild black hole. Furthermore, as we shall see in this paper, it also allows us to find approximate solutions for C-metrics with small $A$ as a perturbation of Schwarzschild solutions.
\par
Geodesics of the C-metric were earlier studied by Pravda and Pravdov\'{a} \cite{Pravda:2000zm, Pravda:2002kj} in C-metric coordinates, as well as Weyl coordinates and coordinates adapted to boost-rotation symmetry. In their paper they considered an analytical solution representing circular orbits around the accelerating black hole and analyzed their stability. It was found that stable timelike circular orbits exist for relatively small values of $mA$ ($\lesssim 4.54\times 10^{-3}$). A corresponding generalization of this work to the anti-de Sitter background was considered in Ref.~\cite{Chamblin:2000mn}. In Ref.~\cite{Podolsky:2003gm}, by using the existence of a conformal Killing tensor in the C-metric, the equations for null geodesics are separated and studied in the context of gravitational radiation in the anti-de Sitter C-metric.
\par
The vacuum C-metric may be viewed as a nonlinear superposition of Schwarzschild and Rindler spacetimes \cite{Bini:2004rr}. Thus, the geodesics of the C-metric should then be physically interpreted as the motion of test particles under the gravitational influence of a black hole, in addition to a uniform constant force. This was denoted as the gravitational Stark effect by Bini et al. \cite{Bini:2004rr, Bini:2004bh}. The \emph{Newtonian} Stark effect\footnote{Also known as the \emph{accelerated Kepler problem} or the \emph{classical Stark effect}.}---which is the motion of particles under a Kepler potential plus a constant force---has been considered in the context of celestial mechanics in Refs.~\cite{Berglund:2000,Namouni:2007vj,Belyaev:2010sm}. The equations of motion under a Newtonian Stark potential are well known to be separable if one uses parabolic coordinates.\footnote{See, e.g., \cite{gignoux2009solved} for a review.} In the case of the C-metric geodesic equations the separability of the Hamilton--
Jacobi equation for timelike particles (if any) is not obvious. We will, however, consider certain special cases where analytical or approximate solutions could be found. 
\par
This paper is organized as follows. In Sec.~\ref{metric} we review the properties of the neutral, Ricci-flat C-metric. The geodesic equations and constants of motion are derived in Sec.~\ref{eom}. In Sec.~\ref{potential} the general, qualitative behavior of the geodesics is considered by studying the effective potential. Sec.~\ref{zerophi} follows by studying the geodesics with zero angular momentum. Here, we shall see that analytical solutions are possible for radial motion along the poles. For non-radial motion, we find approximate solutions describing polar orbits with small $A$. This is possible by considering them as perturbations of circular Schwarzschild orbits of radius $r_0$. We will find that circular timelike orbits with $r_0<6m$ are unstable, while those with $r_0>6m$ will have orbits making small oscillations around $r_0$ when a small $A$ is introduced. In Sec.~\ref{circular} we consider circular orbits encircling the 
acceleration axis and analyze its stability by introducing small displacements to the circular trajectories. A few examples of numerical solutions for bound, co-accelerating orbits are presented in Sec.~\ref{bound}. This paper concludes in Sec.~\ref{conclusion} where a summary and discussion of possible future work is presented.
\par
It is worth noting that the perturbative analysis of Sec.~\ref{zerophi} differs from the one in Sec.~\ref{circular}. The perturbative solutions found in Sec.~\ref{zerophi} correspond to a small $A$ approximation of the geodesic equations, which may be interpreted as orbits around the \emph{perturbed} Schwarzschild spacetime with a small acceleration along the orbital plane. On the other hand, the circular geodesics in Sec.~\ref{circular} are considered for any generic $A$, which need not be small. The perturbation refers to small displacements given to the orbiting particle itself.

\section{The metric} \label{metric}

We consider the vacuum C-metric written in the factorized form, which is given by \cite{Hong:2003gx}
\begin{align}
 \dif s^2\eq\frac{1}{A^2(x-y)^2}\brac{-F(y)\dif\tilde{t}^2+\frac{\dif y^2}{F(y)}+\frac{\dif x^2}{G(x)}+G(x)\dif\phi^2}, \label{general_form}
\end{align}
where
\begin{align}
   F(y)=-\brac{1-y^2}\brac{1+2mAy},\quad G(x)=\brac{1-x^2}\brac{1+2mAx}. \label{Rflat_structures}
\end{align}
The metric is a solution to the vacuum Einstein equations with zero cosmological constant, $R_{\mu\nu}=0$. The parameters $m$ and $A$ are related to the mass and acceleration of the black hole, respectively. For the metric \Eqref{general_form} to have a Lorentzian signature, the coordinates are restricted to $-1<x<1$,  and $-1/2mA<y<-1$, in addition to the condition $2mA\leq 1$. It is often convenient to introduce spherical-type coordinates \cite{Griffiths:2006tk} by the transformation
\begin{align}
 x=\cos\theta,\quad y=-\frac{1}{Ar},\quad \tilde{t}=At.
\end{align}
Then the metric becomes
\begin{align}
 \dif s^2\eq\frac{1}{(1+Ar\cos\theta)^2}\brac{-Q(r)\dif t^2+\frac{\dif r^2}{Q(r)}+\frac{r^2\dif\theta^2}{P(\theta)}+P(\theta)r^2\sin^2\theta\,\dif\phi^2}, \label{Rflat_sph}
\end{align}
where
\begin{align}
 P(\theta)=1+2mA\cos\theta,\quad Q(r)=\brac{1-A^2r^2}\brac{1-\frac{2m}{r}}. \label{Rflat_structures_sph}
\end{align}
The coordinate ranges are now $2m<r<1/A$ and $0\leq\theta\leq \pi$, this only covers the region between the black hole horizon and the acceleration horizon. The periodicity of $\phi$ is $-\pi C_0\leq \phi\leq \pi C_0$, where $C_0$ is chosen to eliminate conical singularities either on the $\theta=0$ or $\theta=\pi$ axis. The singularity at $\theta=0$ can be eliminated by setting $C_0=(1+2Am)^{-1}$. This corresponds to the black hole being accelerated by a cosmic string. On the other hand, setting $C_0=(1-2Am)^{-1}$ eliminates the singularity at $\theta=\pi$, which corresponds to the black hole being pushed by a cosmic strut.

\section{Geodesic equations} \label{eom}

The Lagrangian for geodesic motion in the spacetime \Eqref{general_form} is given by
\begin{align}
 2\mathcal{L}\eq\frac{1}{A^2(x-y)^2}\brac{-F(y)\dot{\tilde{t}}^2+\frac{\dot{y}^2}{F(y)}+\frac{\dot{x}^2}{G(x)}+G(x)\dot{\phi}^2}\eq\epsilon,\label{Lagrangian}
\end{align}
where $\epsilon=-1$ corresponds to timelike geodesics, and $\epsilon=0$ for null geodesics. The geodesic equations can be derived using the Euler--Lagrange equation
\begin{align}
 \frac{\dif}{\dif\tau}\frac{\partial\mathcal{L}}{\partial\dot{q}^\mu}=\frac{\partial\mathcal{L}}{\partial q^\mu},\quad q^\mu(\tau)\equiv\brac{\tilde{t}(\tau),y(\tau),x(\tau),\phi(\tau)}, \label{ELE}
\end{align}
where $\tau$ is an affine parameter that parametrizes the trajectory, and overdots denote derivatives with respect to $\tau$. The two Killing vectors $\partial/\partial t$ and $\partial/\partial\phi$ give rise to two constants of motion $E$ and $\Phi$, which are related to the first integrals of the $\tilde{t}$ and $\phi$ coordinate by
\begin{align}
 \dot{\tilde{t}}=\frac{A^2(x-y)^2\widetilde{E}}{F},\quad \dot{\phi}=\frac{A^2(x-y)^2\Phi}{G}. \label{constants_of_motion_xy}
\end{align}
The constants $\widetilde{E}$ and $\Phi$ can be physically interpreted as the energy and angular momentum of the particle.  Substituting \Eqref{constants_of_motion_xy} into \Eqref{Lagrangian} gives a constraint
\begin{align}
 A^2(x-y)^2\brac{\frac{\Phi^2}{G}-\frac{\widetilde{E}^2}{F}}+\frac{1}{A^2(x-y)^2}\brac{\frac{\dot{y}^2}{F}+\frac{\dot{x}^2}{G}}=\epsilon. \label{first_integral1}
\end{align}
The constraint equation can be rewritten in terms of an effective potential formulation:
\begin{align}
 \frac{F}{A^4(x-y)^4}\brac{\frac{\dot{x}^2}{G}+\frac{\dot{y}^2}{F}}=\widetilde{E}^2-\widetilde{V}_\mathrm{eff}(x,y)^2,
\end{align}
where 
\begin{align}
 \widetilde{V}_\mathrm{eff}(x,y)=\sqrt{\frac{F\Phi^2}{G}-\frac{\epsilon F}{A^2(x-y)^2}}
\end{align}
is the effective potential for the geodesic motion.  We may find the existence of stationary or bound/unbound orbits by analyzing the behavior of $\widetilde{V}_\mathrm{eff}(x,y)$. 
The equations of motion for $x$ and $y$ are obtained by applying Eq.~\Eqref{ELE}, which gives\footnote{The equations of motion here also hold for C-metrics with a nonzero cosmological constant as well (de Sitter or Anti-de Sitter C-metrics, e.g., Refs.~\cite{Dias:2002mi,Dias:2003xp,Krtous:2003tc,Podolsky:2003gm,Krtous:2005ej}) because these metrics have the same form as \Eqref{general_form}, and only differ in the particular forms of the structure functions $F(y)$, $G(y)$ and an overall constant factor.}
\begin{align}
 \ddot{x}=&\brac{\frac{G'}{2G}+\frac{1}{x-y}}\dot{x}^2-\frac{G\dot{y}^2}{(x-y)F}-\frac{2\dot{x}\dot{y}}{x-y}+\nonumber\\
          &\hspace{1.5cm}+A^4(x-y)^3G\sbrac{\frac{\widetilde{E}^2}{F}+\brac{\frac{(x-y)G'}{2G}-1}\frac{\Phi^2}{G}},\label{xddoteqn}\\
 \ddot{y}=&\brac{\frac{F'}{2F}-\frac{1}{x-y}}\dot{y}^2+\frac{F\dot{x}^2}{(x-y)G}+\frac{2\dot{x}\dot{y}}{x-y}\nonumber\\
          &\hspace{1.5cm}-A^4(x-y)^3F\sbrac{\brac{\frac{(x-y)F'}{2F}+1}\frac{\widetilde{E}^2}{F}-\frac{\Phi^2}{G}}. \label{yddoteqn}
\end{align}
Here primes denote derivatives with respect to the argument of the function. ($F'=\dif F/\dif y$ and $G'=\dif G/\dif x$.) The solutions to these equations should also obey Eq.~\Eqref{first_integral1}. This feature can be used as a consistency check for numerics.
\par
In spherical-type coordinates, the conserved quantities are
\begin{align}
 \dot{t}=\frac{(1+Ar\cos\theta)^2}{Q}E,\quad \dot{\phi}=\frac{(1+Ar\cos\theta)^2}{Pr^2\sin^2\theta}\Phi, \label{Rflat_constants_sph}
\end{align}
while the equations of motion are given by
\begin{align}
 \ddot{r}\eq&\brac{\frac{1}{r}-\frac{1}{r(1+Ar\cos\theta)}+\frac{Q'}{2Q}}\dot{r}^2+\frac{rQ\,\dot{\theta}^2}{(1+Ar\cos\theta)P}-\frac{2Ar\sin\theta\,\dot{r}\dot{\theta}}{1+Ar\cos\theta}\nonumber\\
            &-(1+Ar\cos\theta)^3Q\left\{\sbrac{(1+Ar\cos\theta)\brac{\frac{Q'}{2Q}-\frac{1}{r}}+\frac{1}{r}}\frac{E^2}{Q}-\frac{\Phi^2}{Pr^3\sin^2\theta } \right\},\label{rddoteqn}\\
 \ddot{\theta}\eq&\brac{\frac{P'}{2P}-\frac{Ar\sin\theta}{1+Ar\cos\theta}}\dot{\theta}^2+\frac{2AP\sin\theta\,\dot{r}^2}{r(1+Ar\cos\theta)Q}-\frac{2\,\dot{r}\dot{\theta}}{P(1+Ar\cos\theta)}\nonumber\\
            &-(1+Ar\cos\theta)^3P\sin\theta\nonumber\\
            &\times\left\{\frac{AE^2}{rQ}-\sbrac{\frac{1+Ar\cos\theta}{r^4}\brac{\frac{\cos\theta}{\sin^2\theta}+\frac{P'}{2P\sin\theta}}+\frac{A}{r^3}}\frac{\Phi^2}{P\sin^2\theta}\right\}.\label{thetaddoteqn}
\end{align}
The constraint equation in these coordinates is
\begin{align}
 (1+Ar\cos\theta)^2\brac{\frac{\Phi^2}{Pr^2\sin^2\theta}-\frac{E^2}{Q}}+\frac{1}{(1+Ar\cos\theta)^2}\brac{\frac{\dot{r}^2}{Q}+\frac{r^2\,\dot{\theta}^2}{P}}\eq\epsilon. \label{first_integral2}
\end{align}
When cast into the effective potential formulation, the constraint equation becomes
\begin{align}
\frac{1}{(1+Ar\cos\theta)^4}\brac{\dot{r}^2+\frac{r^2Q\,\dot{\theta}^2}{P}}=E^2-V_\mathrm{eff}(r,\theta)^2,
\end{align}
where
\begin{align}
 V_\mathrm{eff}(r,\theta)=\sqrt{Q\brac{\frac{\Phi^2}{Pr^2\sin^2\theta}-\frac{\epsilon}{(1+Ar\cos\theta)^2}}}. \label{V_eff}
\end{align}
Eqs.~\Eqref{rddoteqn}, \Eqref{thetaddoteqn}, and \Eqref{Rflat_constants_sph} can be solved numerically using the fourth-order Runge Kutta algorithm. Recalling that the periodicity of $\phi$ is $-\pi C_0\leq \phi\leq \pi C_0$, the results may be visualized in Cartesian-like coordinates with
\begin{align}
 X=r\sin\theta\cos\frac{\phi}{C_0},\quad Y=r\sin\theta\sin\frac{\phi}{C_0},\quad Z=r\cos\theta.
\end{align}

\section{Effective potential} \label{potential}

\subsection{Timelike geodesics}
\begin{figure}[ht]
  \begin{center}
    \includegraphics[scale=0.75]{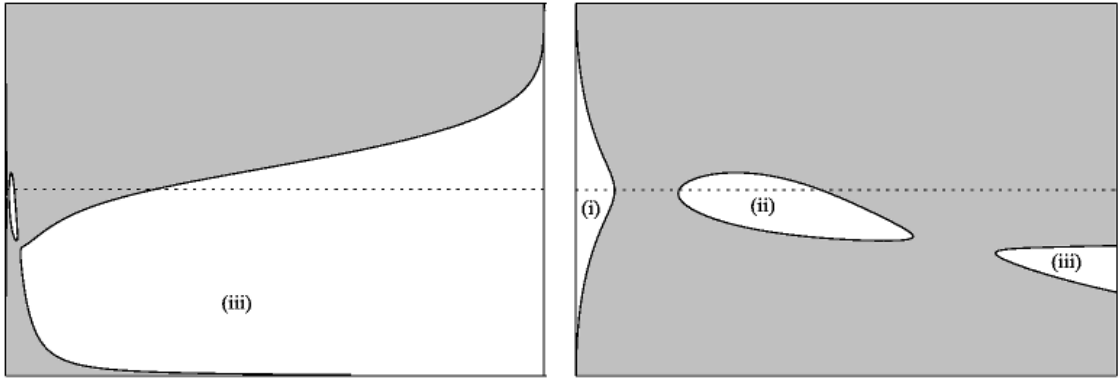}
    \caption{A typical shape of an equipotential curves of $V_\mathrm{eff}^2=\mathrm{const}.$ for timelike particles. The shaded areas are regions not accessible to the particles. The vertical axis represents the $\theta$-coordinate and the horizontal axis is for $r$, where the left edge of each plot corresponds to the horizon $r=2m$. The left figure shows the equipotential lines over the entire range $2m<r<1/A$, where regions (i) and (ii) are too small to be visible. These regions are shown more clearly in the right plot focusing on the range of $r$ close to $2m$. The dotted horizontal line shows the equator $\theta=\pi/2$.}
    \label{czv_regions}
  \end{center}
\end{figure}

Before proceeding to solve the equations of motion, we first make some general observations regarding the behavior of the geodesics by studying the effective potential in Eq.~\Eqref{V_eff}.  The typical shapes of equipotential curves of $V_\mathrm{eff}^2=\mathrm{constant}$ for timelike particles are shown in Fig.~\ref{czv_regions}. In this figure, we see that the geodesics can typically reside in one of three regions, corresponding to three possible outcomes for the particles, namely (i) falling into the black hole horizon $r=2m$, (ii) bound geodesics co-accelerating with the black hole, and (iii) falling into the acceleration horizon $r=1/A$. Typically regions (i) and (ii) are small relative to the full coordinate range $2m<r<1/A$, (they are barely visible in the left plot of Fig.~\ref{czv_regions}). The right plot of Fig.~\ref{czv_regions} focuses on a range of $r$ close to the black hole such that (i) and (ii) are visible.
\par
It is interesting to observe that a large portion of region (iii) lies under the $\theta=\pi/2$ line, which brings particles closer to the $\theta=0$ pole. Recalling that the black hole is accelerated by the cosmic string/strut in the $\theta=\pi$ direction, we have the intuitive notion that the particles tend to get `left behind' by the accelerating black hole.

\par
\begin{figure}
 \begin{center}
  \includegraphics[scale=0.75]{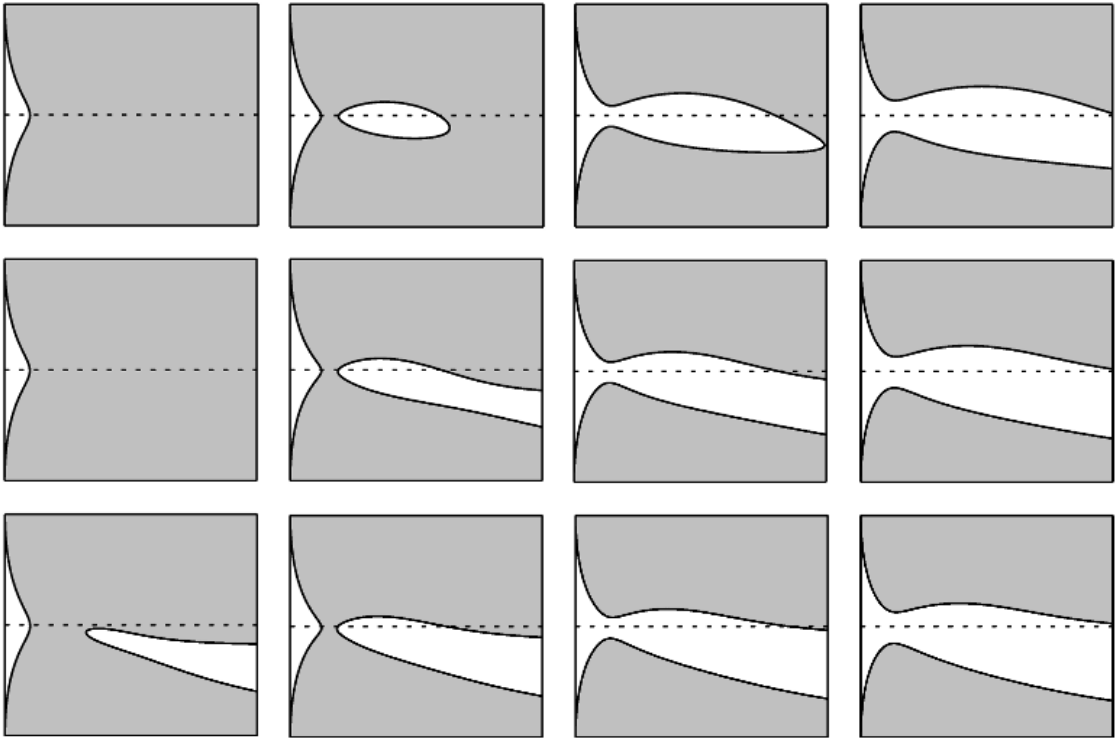}
  \caption{Plots of $V_\mathrm{eff}^2=E^2$ for timelike geodesics at fixed angular momentum $\Phi^2=13$ and various $A$ and $E$. The first, second and third rows respectively are plots for $A=0.001$, 0.002 and 0.003. The columns from left to right are plots for $E^2=0.90$, 0.91, 0.92 and 0.93. The shaded regions indicate areas not accessible to the particles. The horizontal and vertical axes are $r$ and $\theta$ respectively.}
  \label{czvplots_timelike}
 \end{center}
\end{figure}

In general, the regions (i), (ii) and (iii) are not always distinct from each other. By tuning the values of $\Phi$, $E$ and $A$, it is possible that region (i) becomes connected to (ii), region (ii) connects to (iii), or all three regions connecting together such that the particles have possible access to the black hole or the acceleration horizon. (See Fig.~\ref{czvplots_timelike}.)

\subsection{Null geodesics}

\begin{figure}
 \begin{center}
  \includegraphics[scale=0.75]{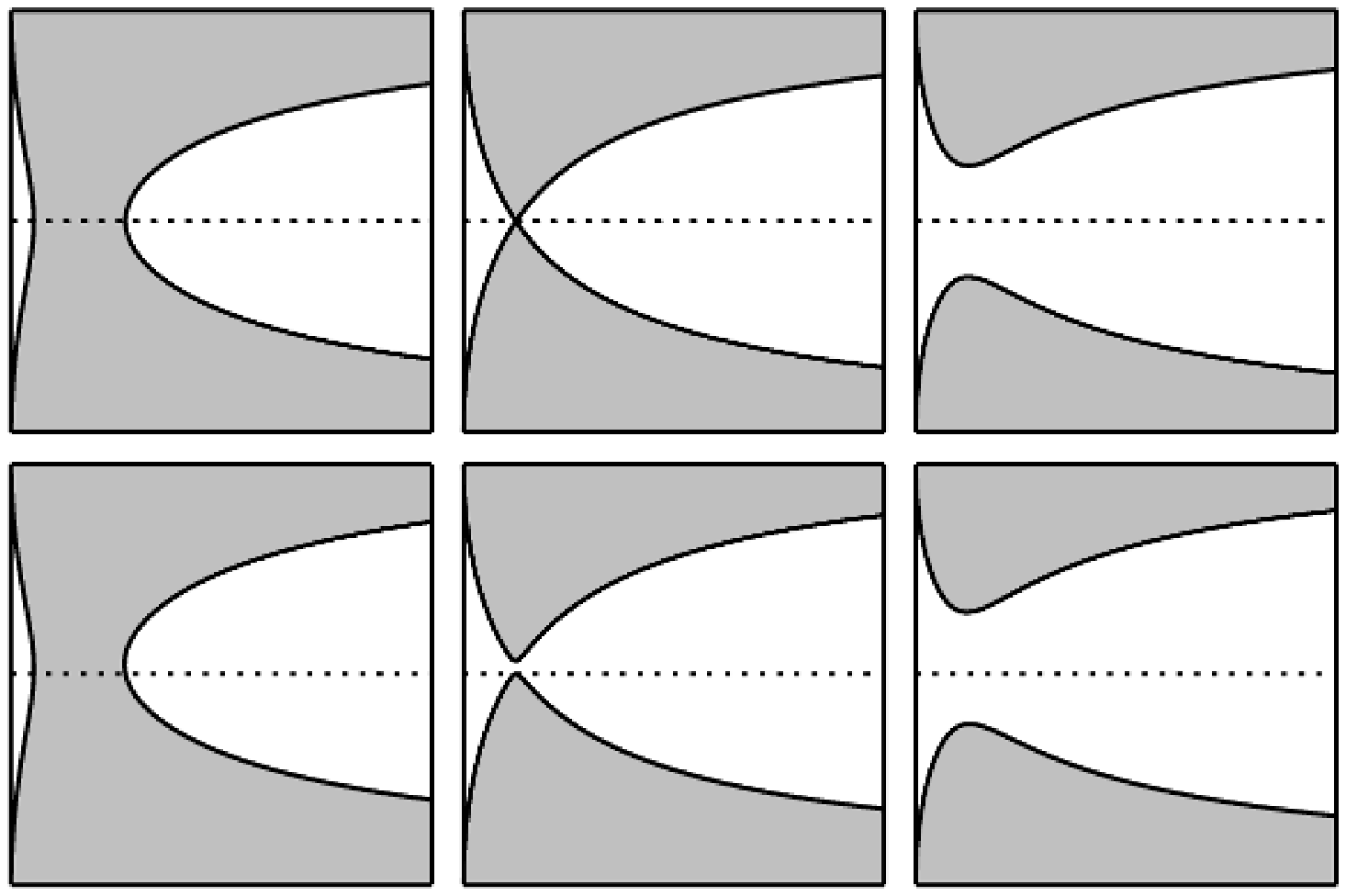}
  \caption{Plots of $V_\mathrm{eff}^2=E^2$ for null geodesics at fixed angular momentum $\Phi^2=14$, $m=1$ and various $A$ and $E$. The first and second rows respectively, are plots for $A=0$, and 0.01. The columns from left to right are plots for $E^2=0.4185$, $0.5185$, and $0.6185$. The middle value of $E$ is chosen to be $E^2=\Phi^2/27m$ to correspond to energies for which spherical photon orbits are possible in the Schwarzschild ($A=0$) case. The shaded regions indicate areas not accessible to the particles. The horizontal and vertical axes are $r$ and $\theta$ respectively.}
  \label{czvplots_null}
 \end{center}
\end{figure}

For null geodesics, since $\epsilon=0$, it can be seen from Eq.~\Eqref{first_integral2} that the equation $\dot{r}=\dot{\theta}=0$ has three roots for $r$ at most (one of them is possibly negative). Therefore curves of $V^2_\mathrm{eff}$ can at most separate into two distinct regions accessible by null geodesics. (See Fig.~\ref{czvplots_null}.) Stable, bound co-accelerating orbits are not possible, consistent with the Schwarzschild case, with the particles either falling into the black hole horizon or through the acceleration horizon. Circular photon orbits of constant $r$ and $\theta$ are possible, and it was shown analytically in \cite{Pravda:2000zm} that such orbits are unstable. This is analogous to the unstable Schwarzschild circular photon orbits of $r=3m$.

\section{Geodesics with zero angular momentum} \label{zerophi}

We first consider geodesics with $\Phi=0$. In this case the equations of motion remain finite at the axes where $\theta=0$ or $\pi$. Therefore we can consider geodesics passing through the poles. Strictly speaking, geodesics that pass trough \emph{both} $\theta=0$ and $\theta=\pi$ are not possible because the particles would collide with the cosmic string or strut responsible for the black hole acceleration. We can however assume that it is indeed possible for the particles to pass arbitrarily close to the axis, and hence the geodesic equations at $\theta=0$, $\pi$ holds as an approximation.

\subsection{Radial geodesics along the poles}

\subsubsection{Timelike radial geodesics along the north pole}

Along the north pole ($\theta=0$), Eq.~\Eqref{V_eff} reduces to
\begin{align}
 V_\mathrm{eff}(r)^2\eq\frac{(r-2m)(1-Ar)}{r(1+Ar)}. \label{V2north}
\end{align}
A typical plot of $V^2_\mathrm{eff}$ is shown in Fig.~\ref{polarnorth}. We see that there exists an unstable equilibrium point at
\begin{align}
 r\eq\frac{mA+\sqrt{2m^2A^2+mA}}{A(1+mA)}
\end{align}
where $\dif\brac{V^2_\mathrm{eff}}/\dif r=0$. Therefore it is possible for a radial timelike particle to be co-accelerated at a fixed distance `behind' the accelerating black hole. However this eqilibrium is unstable as it can be shown that $\dif^2\brac{V^2_\mathrm{eff}}/\dif r^2$ is negative. A small perturbation will cause the particle to either fall into the black hole, or fall behind the acceleration horizon.
\par
At the north pole Eq.~\Eqref{first_integral2} becomes
\begin{align}
 \dot{r}\eq\pm\sbrac{(1+Ar)^{3/2}}\sqrt{\frac{A\brac{1+E^2}r^2-\brac{1-E^2+2mA}r+2m}{r}}. \label{rdot_polar_north}
\end{align}
The turning points of the particle motion are found by solving $\dot{r}=0$, where we find
\begin{align}
 r_\pm\eq\frac{1-E^2+2mA\pm\sqrt{\brac{1-E^2}^2-4mA\brac{3E^2+1}r+4m^2A^2}}{2A(1+E^2)}.
\end{align}
At the north pole, $r_+>r_->0$. The regions accessible by the particles are either $2m< r\leq r_-$, or $r\geq r_+$; for a given energy $E$, there is a potential barrier at $r_-<r<r_+$. As already seen by the analysis of $V^2_\mathrm{eff}$ above, a value of $E$ may be chosen such that the two roots coincide, and the potential barrier vanishes. The particle sitting at $r=r_-=r_+$ will be at the point of unstable equilibrium.
\par
We can rewrite Eq.~\Eqref{rdot_polar_north} in terms of the roots $r_\pm$ and solve it for a timelike particle starting at an initial position $r_\mathrm{init}$ and obtain
\begin{align}
 \tau(r)\eq\int_{r_\mathrm{init}}^{r}\frac{\sqrt{r'}\;\dif r'}{\sqrt{A(1+E^2)(1+Ar')^3(r_--r')(r_+-r')}},
\end{align}
where we may set $\tau(r_\mathrm{init})=0$. The above integration can be expressed in terms of various elliptic integrals. However the expressions are long and cumbersome, and not particularly illuminating; so we will not show them here. For specific values of $E$ and $A$ this may be calculated easily with the use of a computer algebra package. As an example, for $A=0.005$ and $m=1$, a particle with energy $E^2=0.7$ will see a potential barrier in the region $r_-<r<r_+$. If it starts from rest at $r=r_+$, is separated from the black hole horizon by a potential barrier. It will fall beyond the acceleration horizon in proper time $\tau\brac{\frac{1}{A}}\simeq193.46$. On the other hand, if it starts from rest at $r=r_-$ instead, it falls in the potential well toward the black hole in proper time $\tau\brac{2m}\simeq27.782$.

\subsubsection{Timelike radial geodesics along the south pole}

\begin{figure}[H]
\begin{center}
 \begin{subfigure}[b]{0.4\textwidth}
  \centering
  \includegraphics[scale=1]{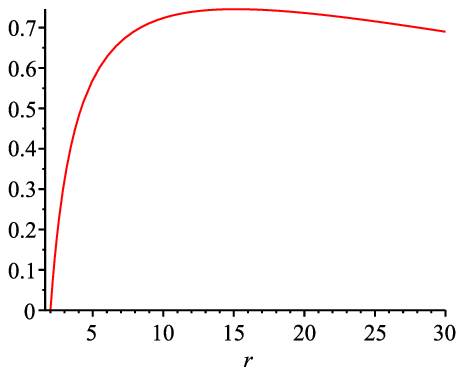}
  \caption{$\theta=0$}
  \label{polarnorth}
 \end{subfigure}
 \hspace{0.2cm}
 \begin{subfigure}[b]{0.4\textwidth}
  \centering
  \includegraphics[scale=1]{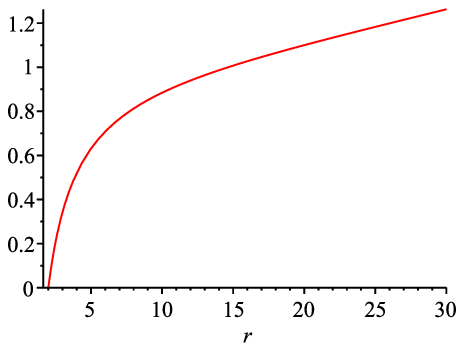}
  \caption{$\theta=\pi$}
  \label{polarsouth}
 \end{subfigure}
\end{center}
\caption{Plots of radial potential $V_\mathrm{eff}^2$ with zero angular momentum at (a) $\theta=0$ and (b) $\theta=\pi$. The values used for this figure are $m=1$, $A=0.005$.}
\label{polar}
\end{figure}

Along the south pole ($\theta=\pi$), Eq.~\Eqref{V_eff} reduces to
\begin{align}
 V^2_\mathrm{eff}(r)\eq\frac{(r-2m)(1+Ar)}{r(1-Ar)}.
\end{align}
A plot of this potential is shown in Fig.~\ref{polarsouth}. The equations for the case along the south pole can simply be obtained by replacing $A\rightarrow -A$ in Eq.~\Eqref{V2north}. For this case, there is no value of $r>0$ that satisfies $\dif\brac{V^2_\mathrm{eff}}/\dif r=0$. Therefore there are no equilibrium points for particles along the south pole; it is not possible for a timelike particle to be co-accelerated at a fixed distance `ahead' of the black hole.
\par
Turning to Eq.~\Eqref{first_integral2}, the equation of motion is
\begin{align}
 \dot{r}\eq\pm(1-Ar)^{3/2}\sqrt{\frac{-A\brac{1+E^2}r^2-\brac{1-E^2-2mA}r+2m}{r}}. \label{rdot_polar_south}
\end{align}
we find turning points of the particle motion are found by solving $\dot{r}=0$, where we get
\begin{align}
 r_\pm\eq\frac{-\brac{1-E^2}+2mA\pm\sqrt{\brac{1-E^2}^2+4mA\brac{3E^2+1}r+4m^2A^2}}{2A\brac{E^2+1}},
\end{align}

Here at the south pole, only the upper root $r_+$ is positive. Therefore the range of allowed motion is $2m<r\leq r_+$. We can rewrite Eq.~\Eqref{rdot_polar_south} in terms of the roots $r_\pm$ and solve it for a timelike particle starting at an initial position $r_\mathrm{init}$ and obtain
\begin{align}
 \tau(r)\eq\int_{r_\mathrm{init}}^{r}\frac{\sqrt{r'}\;\dif r'}{\sqrt{A(1+E^2)(1-Ar')^3(r'-r_-)(r_+-r')}},
\end{align}
where we have set $\tau(r_\mathrm{init})=0$. As an example, for $A=0.02$ and $m=1$, a particle with energy $E^2=0.9$ initially at rest at $r=r_+$, the proper time for it to fall into the horizon is given by $\tau(2m)\simeq15.647$.

\subsubsection{Radial null geodesics}

The analysis of radial null geodesics is much simpler since for $\epsilon=0$, in addition to $\Phi=0$, Eq.~\Eqref{first_integral2} simplifies into
\begin{align}
 \dot{r}\eq\pm E\brac{1+Ar\xi}^2,
\end{align}
where $\xi\eq\cos\theta$. (Therefore $\xi=1$ at the north pole, and $\xi=-1$ at the south.) The equation is solved by
\begin{align}
 \pm E\tau(r)\eq\frac{1}{A\xi(1+Ar\xi)}+\mathrm{const}. \label{tau_soln_null}
\end{align}
If we consider the $\dot{t}$ equation \Eqref{Rflat_constants_sph} to obtain
\begin{align}
 \frac{\dif r}{\dif t}\eq \pm Q.
\end{align}
The solution is
\begin{align}
 \pm t(r)\eq\frac{2m\ln(r-2m)}{1-4m^2A^2}+\frac{\ln(1+Ar)}{2A(1+2mA)}-\frac{\ln(1-Ar)}{2A(1-2mA)}+\mathrm{const}.\label{t_soln_null}
\end{align}
We see from Eq.~\Eqref{tau_soln_null} that radial photons fall into the horizon in finite proper time, but takes infinite coordinate time to reach the black hole horizon or the acceleration horizon.

\subsection{Polar orbits around weakly accelerated black holes}

\subsubsection{Perturbations around timelike Schwarzschild geodesics}

For trajectories with $\dot{\theta}\neq 0$, and constant $\phi$, we extend the coordinate range of $\theta$ to cover the range $\theta\in [0,2\pi]$ to describe the regions accessible by a polar orbit. When $\theta$ is non-constant, the equations of motion for $\dot{\theta}\neq 0$ are difficult to solve. However, a possible case we can study is for small acceleration $A$, treated as a perturbation of circular Schwarzschild geodesics. For the case $A=0$, the equations of motion reduce to that of Schwarzschild. We denote circular Schwarzschild orbits of energy $E$ as having constant radius $r=r_0$, and linearly increasing $\theta_0=L\tau/r_0^2$, where $L$ is the angular momentum in the $\theta$ direction. Expressed in terms of $r_0$ and $m$, the energy and angular momentum are \cite{Chandrasekhar:1985kt,Carroll:2004st}
\begin{align}
 L^2\eq\frac{r_0^2 m}{r_0-3m},\quad E^2\eq\frac{(r_0-3m)^2}{r_0(r_0-3m)}. \label{Sch_circular_EL}
\end{align}
Considering small $A$, we write
\begin{align}
 r(\tau)\eq& r_0+Ar_1(\tau)+\mathcal{O}\brac{A^2},\quad \theta(\tau)\eq\frac{L}{r_0^2}\tau+A\theta_1(\tau)+\mathcal{O}\brac{A^2},
\end{align}
and substitute into Eqs.~\Eqref{rddoteqn} and \Eqref{thetaddoteqn}. When expanded to linear order in $A$ the equations become
\begin{align}
 \ddot{r}_1\eq&\frac{3r_0-2m}{r_0^3(r_0-3m)}r_1+\frac{2(r_0-2m)}{r_0\sqrt{r_0-3m}}\dot{\theta}_1\nonumber\\
              &\quad+\frac{(r_0^2-7mr_0-2m^2)(r_0-2m)}{r_0^2(r_0-3m)}\cos\brac{\frac{\sqrt{m}}{r_0\sqrt{r_0-3m}}\tau}, \label{polarr1ddot}\\
 \ddot{\theta}_1\eq&-\frac{2\sqrt{m}}{r_0^2\sqrt{r_0-3m}}\dot{r}_1-\frac{r_0^2-mr_0+m^2}{r_0^2(r_0-3m)}\sin\brac{\frac{\sqrt{m}}{r_0\sqrt{r_0-3m}}\tau}. \label{polartheta1ddot}
\end{align}
In the above, Eq.~\Eqref{Sch_circular_EL} has been used to express $E$, and $L$ in terms of $m$ and $r_0$. Eq.~\Eqref{polartheta1ddot} can be integrated once directly; the resulting $\dot{\theta}_1$ is
\begin{align}
 \dot{\theta}_1\eq&-\frac{2\sqrt{m}}{r_0^2\sqrt{r_0-3m}}r_1+\frac{\brac{m^2-mr_0+r_0^2}}{r_0\sqrt{m(r_0-3m)}}\cos\brac{\frac{\sqrt{m}}{r_0\sqrt{r_0-3m}}\tau}+K, \label{theta1dotsub}
\end{align}
where $K$ is the integration constant, which may be fixed by considering the following initial conditions:
\begin{align}
 r_1(0)\eq\dot{r}_1(0)\eq0\eq\theta_1(0). \label{polar_perturb_init}
\end{align}
By the using the same initial conditions on Eq.~\Eqref{first_integral2} and comparing with Eq.~\Eqref{theta1dotsub}, the integration constant is fixed as $K=-m^{3/2}/r_0\sqrt{r_0-3m}$. The resulting expression for $\dot{\theta}_1$ may be substituted into \Eqref{polarr1ddot} to give
\begin{align}
 \ddot{r}_1\eq&-\omega^2r_1+\frac{3(r_0-2m)}{r_0}\cos\Omega\tau-\frac{2(r_0-2m)m^2}{r_0^2(r_0-3m)}, \label{r1ddot_final}
\end{align}
where
\begin{align}
 \omega\eq\sqrt{\frac{(r_0-6m)m}{(r_0-3m)r_0^3}},\quad \Omega\eq\frac{\sqrt{m}}{r_0\sqrt{r_0-3m}}. \label{frequencies}
\end{align}
We see in Eq.~\Eqref{r1ddot_final} that $\omega^2<0$ for $r_0<6m$, signalling an instability of the orbit. (The solution for $r_1$ in this case would be exponentially growing in magnitude.) On the other hand, for $r_0>6m$, we have $\omega^2>0$ and Eq.~\Eqref{r1ddot_final} takes the form a harmonic oscillator of frequency $\omega$, driven by a periodic force of frequency $\Omega$, in addition to a constant force. Resonance is not possible, since from Eq.~\Eqref{frequencies} we have
\begin{align}
 \frac{\omega}{\Omega}\eq\sqrt{1-\frac{6m}{r_0}}.
\end{align}
Therefore $\omega\eq\Omega$ only for $r_0\rightarrow\infty$. We conclude that perturbed orbits with $r_0>6m$ are stable under a weak acceleration of the black hole.
\par
\begin{figure}
 \begin{center}
  \includegraphics[scale=1]{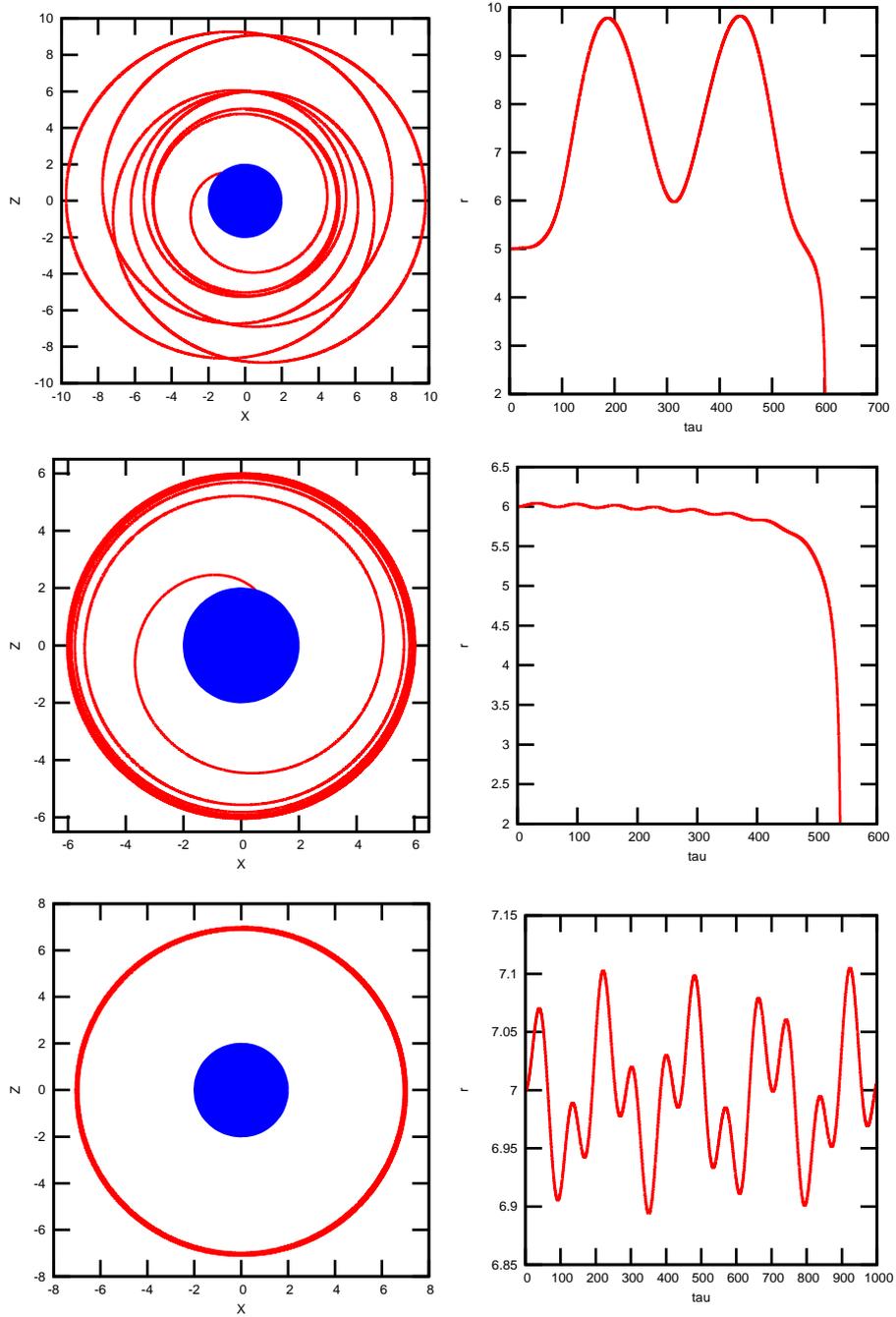}
  \caption{Perturbation of Schwarzschild circular orbits under acceleration along the orbital plane. The figures on the left side are geodesics plotted in $(X,Z)$ coordinates, while the figures on the right shows $r$ vs $\tau$. All geodesics here are with $A=0.0001$ and $m=1$. From top to bottom, the unperturbed circular orbits are $r_0=5m$, $6m$ and $7m$.}
  \label{perturbpolar}
 \end{center}
\end{figure}

The solution to Eq.~\Eqref{r1ddot_final}, for $r_0>6m$ satisfying the initial conditions \Eqref{polar_perturb_init} is
\begin{align}
 r_1(\tau)\eq&\frac{r_0(r_0-2m)\brac{r_0^3-9mr_0^2+18r_0m^2+4m^3}}{2m^2(r_0-6m)}\cos\omega\tau+\frac{(r_0-2m)(r_0-3m)r_0^2}{2m}\cos\Omega\tau\nonumber\\
          &-\frac{2mr_0(r_0-2m)}{r_0-6m},  \label{r1_soln}
\end{align}
and the solution to $\theta_1$ is
\begin{align}
 \theta_1(\tau)\eq&\frac{(r_0-2m)\brac{r_0^3-9mr_0^2+18r_0m^2+4m^3}\sqrt{r_0}}{m^2\brac{r_0-6m}^{3/2}}\sin\omega\tau\nonumber\\
                  &+\frac{m^3+5r_0m^2-4mr_0^2+r_0^3}{m^2}\sin\Omega\tau + \frac{m^{3/2}\brac{3r_0-2m}}{r_0\sqrt{r_0-3m}(r_0-6m)}\tau.\label{theta1_soln}
\end{align}
These results can be checked with the numerical solution, as shown in Fig.~\ref{perturbpolar}. It can be seen that the bottom figure with $r_0>6m$ exhibits a stable oscillation with its natural frequency $\omega$ and driving frequency $\Omega$. The other top figure has an exponentially decaying solution, while the middle row shows the critical solution where $\omega=0$.
\par
It should be noted that the last term of Eq.~\Eqref{theta1_soln} increases linearly in $\tau$; hence, $\theta_1$ increases linearly as $\tau$ progresses, eventually taking it beyond the validity of the linearized equations \Eqref{polarr1ddot} and \Eqref{polartheta1ddot}. Therefore, the first-order solution will eventually lose its accuracy as the evolution continues.

\subsubsection{Perturbations around null Schwarzschild geodesics}
Circular null geodesics exist around Schwarzschild black holes, but are unstable. Therefore we may expect that perturbing the spacetime with small accelerations should not yield any stable oscillations around the original circular orbits. To see this explicitly, we write
\begin{align}
 r(\tau)\eq 3m+Ar_1(\tau)+\mathcal{O}\brac{A^2},\quad \theta(\tau)\eq\frac{L}{9m^2}\tau+A\theta_1(\tau)+\mathcal{O}\brac{A^2}
\end{align}
and substitute into Eqs.~\Eqref{rddoteqn} and \Eqref{thetaddoteqn} and expand to linear order in $A$. Similar to the timelike case, the equation for $\theta_1$ can be directly integrated to yield an expression for $\dot{\theta}$ with an integration constant. Again this integration constant is fixed by comparing the expansion of \Eqref{first_integral2} (this time with $\epsilon=0$). The result is substituted into the equation of motion for $r_1$, giving
\begin{align}
 \ddot{r}_1\eq\frac{L^2}{81m^4}r_1-\frac{2L^2\brac{7+12L}}{81m^2}.
\end{align}
The coefficient of $r_1$ is always positive, indicating an instability. This result is perhaps not surprising since circular photon orbits around a Schwarzschild black hole ($A=0$) are already unstable.

\subsection{Polar orbits for general \texorpdfstring{$A$}{A}}

\subsubsection{Timelike particles}

\begin{figure}
\begin{center}
 \begin{subfigure}[b]{0.4\textwidth}
  \centering
  \includegraphics[scale=1]{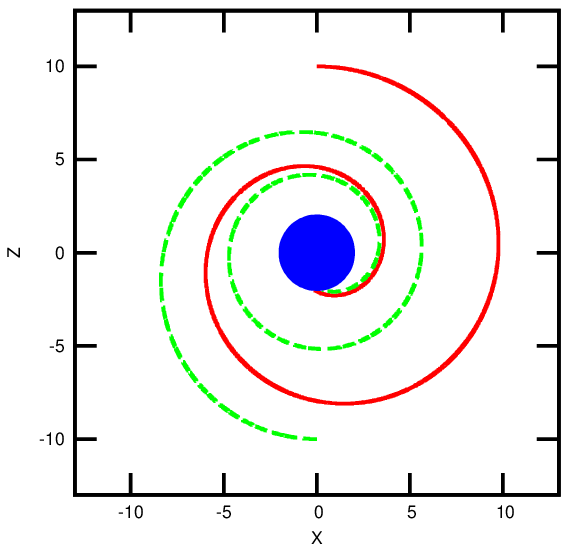}
  \caption{$E^2=0.90$}
  \label{polarE=0.90}
 \end{subfigure}
 %\hspace{0.2cm}
 \begin{subfigure}[b]{0.4\textwidth}
  \centering
  \includegraphics[scale=1]{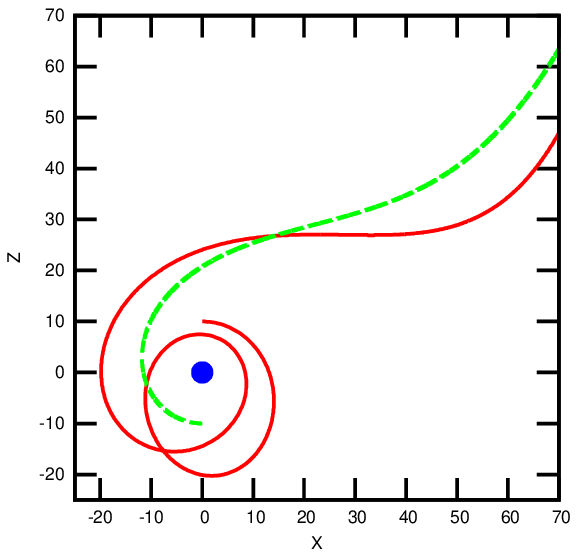}
  \caption{$E^2=0.95$}
  \label{polarE=0.95}
 \end{subfigure}
\end{center}
\caption{Timelike trajectories starting from $\theta=0$ or $\theta=\pi$ passing through the poles with $\Phi=0$ and $A=0.0005$. For each case the initial radial coordinate is $r=10$. The trajectories in Figure (a) are those with $E^2=0.90$, fall into the horizon, while particles in (b) escape the black hole and fall beyond the acceleration horizon. The solid lines represent particles with initial condition $\theta=0$, while the dashed lines are particles starting from $\theta=\pi$.}
\label{polarorbits}
\end{figure}

For general values of $A$ that are not necessarily small, we study the solutions numerically. Typical trajectories are shown in Fig.~\ref{polarorbits}. Timelike particles circling the black hole in the polar direction either fall into the black hole (Fig.~\ref{polarE=0.90}), or get left behind by the accelerating black hole (Fig.~\ref{polarE=0.95}). (Recall that the direction of acceleration in the C-metric is at $\theta=\pi$.)
\par
\begin{figure}
  \begin{center}
   \begin{subfigure}[b]{0.35\textwidth}
    \includegraphics[scale=0.9]{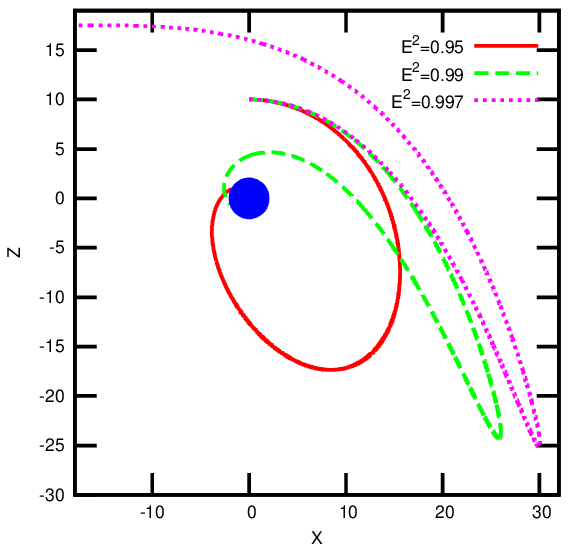}
   \caption{$E^2$ from 0.9 to 0.997.}
   \label{A0.001polara} 
   \end{subfigure}
   \begin{subfigure}[b]{0.35\textwidth}
    \includegraphics[scale=0.9]{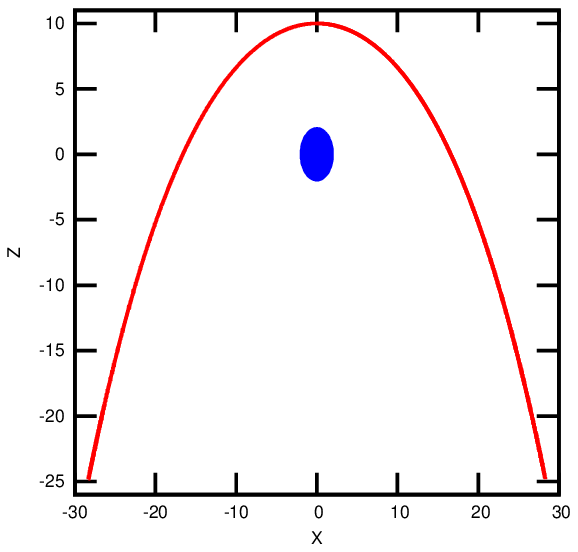}
    \caption{$E^2\simeq 0.994243$.}
    \label{A0.001polarb}
   \end{subfigure}
   \caption{Polar orbits starting from $r=10$, $A=0.001$, with different values of $E$. Geodesics are plotted in the constant $\phi$ plane; therefore, the black hole is accelerated downward ($\theta=\mathrm{const}$) in this figure. It can be seen in Fig.~\ref{A0.001polara} that (i) orbits of $E^2=0.99$ and below will fall into the black hole, and (ii) orbits of $E^2=0.997$ and above escape and eventually pass the acceleration horizon. In the critical point between cases (i) and (ii) the particle oscillates back and forth on the curve shown in Fig.~\ref{A0.001polarb}. The value of energy is found by fine-tuning to obtain $E^2\simeq 0.994243$.}   
   \label{A0.001polar}
  \end{center}
 \end{figure}

If we fix the starting position and consider the outcome of geodesics of different energies, we have three possibilities (i) particles falling into the black hole horizon, (ii) particles escaping to the acceleration horizon, and (iii) a critical point neither falling into the black hole nor escaping to the acceleration horizon. This latter case may be considered as a limiting case between (i) and (ii), and is shown in Fig.~\ref{A0.001polarb}, where the numerical solution starts at $\theta=0$ and it appears to oscillate back and forth between two turning points. This solution was found by fine tuning, as energies slightly higher or lower will respectively, result in the particle falling into the black hole or beyond the acceleration horizon. 

\section{Stability of circular orbits} \label{circular}

In this section, we obtain solutions corresponding to circular orbits in which the orbital plane is perpendicular to the axis of acceleration. Circular orbits are those with constant $r$ and $\theta$. When these values are constant, Eq.~\Eqref{Rflat_constants_sph} implies that $\phi$ increases linearly in $\tau$. Orbits of this kind trace out a circle in a plane of constant $\theta$. Such orbits were considered in \cite{Pravda:2000zm} and were generalized to the case of anti-de SItter C-metrics in Ref.~\cite{Chamblin:2000mn}, which we briefly review in the following.

\subsection{Circular orbit solution}

It will be more convenient in this section to revert to the C-metric coordinates $(x,y)$, which are also constant for circular orbits. Hence we denote the $x(\tau)=x_0$ and $y(\tau)=y_0$ as the solutions corresponding to circular orbits. We can find these solutions analytically by setting $\ddot{x}=\dot{x}=0=\ddot{y}=\dot{y}$ in Eqs.~\Eqref{xddoteqn} and \Eqref{yddoteqn}, giving
\begin{align}
 \frac{E^2}{F}+\brac{\frac{(x-y)G'}{2G}-1}\eq&0,\label{eg1}\\
 \brac{\frac{(x-y)F'}{2F}+1}\frac{E^2}{F}-\frac{\Phi^2}{G}\eq&0. \label{eg2}
\end{align}
Eliminating $\Phi$ and $E$ from Eqs.~\Eqref{eg1} and \Eqref{eg2} leaves us with
\begin{align}
 m^2A^2\brac{3x_0^2y_0^2-x_0^2-y_0^2-4x_0y_0-1}+mA(x_0+y_0)(x_0y_0-3)-1=0. \label{circ_xy}
\end{align}
Particles lying on points on the curve defined by Eq.~\Eqref{circ_xy} with initial conditions $\dot{x}=\dot{y}=0$ will be in circular orbit around the black hole. Using either Eq.~\Eqref{eg1} or \Eqref{eg2} in the first integral \Eqref{first_integral1} gives the corresponding energies and angular momenta of circular orbits
\begin{align}
 E_0^2=\frac{2\epsilon F(y_0)^2}{A^2(x_0-y_0)^3F'(y_0)},\quad \Phi_0^2=\frac{2\epsilon G(x_0)^2}{A^2(x_0-y_0)^3G'(x_0)}. \label{EPhi0_xy}
\end{align}

\subsection{Perturbations of circular orbits}
In Refs.~\cite{Pravda:2000zm} and \cite{Chamblin:2000mn}, the stability of circular orbits was investigated by looking for the existence of local minima in the potential $V_\mathrm{eff}$.  In this paper, we try a different, but equivalent approach by perturbing about the circular geodesics $(x_0,y_0)$ and checking to see if the eigenfrequecies of the perturbed geodesic equations are real.
\par
\begin{figure}[ht]
\begin{center}
 \includegraphics[scale=0.4]{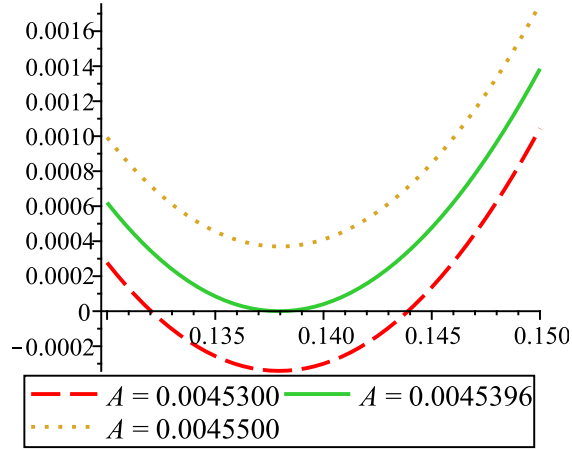}
 \caption{Plot of $\lambda_+$ vs $x_0$. We see that for curves with $A<A_\mathrm{crit}$, there exists range of $x_0$ with negative $\lambda_+$, thus having a stable circular orbit. The critical point $A=A_\mathrm{crit}$ is given numerically in \Eqref{A_crit}, which agrees with the results of \cite{Pravda:2000zm}.}
 \label{lambdaplot}
\end{center}
\end{figure}

We introduce the perturbations by writing
\begin{align}
 x(\tau)\eq& x_0+x_1(\tau)\varepsilon+\mathcal{O}\brac{\varepsilon^2},\quad y(\tau)\eq y_0+y_1(\tau)\varepsilon+\mathcal{O}\brac{\varepsilon^2}. \label{xy_perturb}
\end{align}
We substitute Eqs.~\Eqref{xy_perturb} and \Eqref{EPhi0_xy} into the equations of motion \Eqref{xddoteqn} and \Eqref{yddoteqn}, and expand up to first order in $\varepsilon$. This gives a linear coupled system
\begin{align}
 \frac{\dif^2}{\dif\tau^2}\left(
  \begin{array}{c}
   x_1 \\
   y_1
  \end{array}
 \right) =
 \left(
  \begin{array}{cc}
   W(x_0,y_0) & B(x_0,y_0) \\
   B(x_0,y_0) & \overbar{W}(x_0,y_0)
  \end{array}
 \right)
 \left(
  \begin{array}{c}
   x_1 \\
   y_1
  \end{array}
 \right),
\end{align}
where we have defined
\begin{align}
B(x_0,y_0)\eq&\fracb{3}{A^2(x_0-y_0)^3},\nonumber\\
 W(x_0,y_0)\eq&\frac{1}{(1-x_0)^2(mA-x_0-3mAx_0^2)(x_0-y_0)^3(1+2mAx_0)A^2}\nonumber\\
    &\times\Bigl[6m^2A^2x_0^5-mA(24mAy_0-1)x_0^4+4mA(3mA-y_0)x_0^3\nonumber\\
    &\quad\quad+(12m^2A^2y_0-3y_0+18mA)x_0^2+2(2-m^2A^2)x_0\nonumber\\
    &\quad\quad-y_0-4m^2A^2y_0-3mA\Bigr],\nonumber \\
\overbar{W}(x_0,y_0)\eq&W(y_0,x_0).
\end{align}
The characteristic equation for the system is
\begin{align}
 \lambda^2-\brac{W+\overbar{W}}\lambda+W\overbar{W}-B^2=0, \label{characteristic_polynomial}
\end{align}
which is solved by
\begin{align}
 \lambda_\pm=\half\brac{W+\overbar{W}}\pm\half\sqrt{\brac{W+\overbar{W}}^2+4\brac{B^2-W\overbar{W}}}. \label{eigen_soln}
\end{align}
The stability of the orbits depends on whether both normal mode frequencies $\omega=\sqrt{-\lambda}$ are real, or, in other words, both eigenvalues $\lambda_\pm$ must be negative. This is guaranteed if the larger one is negative, i.e., $\lambda_+>0$. Setting $m=1$, we obtain a plot of $\lambda_+$ vs $x_0$ in Fig.~\ref{lambdaplot}. In the figure, we see that as $A$ increases, the range of $x_0$ having stable circular orbits (negative $\lambda_+$) decreases. If $A$ exceeds a certain critical value $A=A_\mathrm{crit}$, then $\lambda_+>0$ for all $x_0$, and there are no more stable circular orbits.

If we examine Eq.~\Eqref{eigen_soln}, we see that $A_\mathrm{crit}$ corresponds to $\lambda_+=0$ having one real root for $x_0$ within the range $-1<x_0<1$. This occurs when the term under the square root in \Eqref{eigen_soln} equals $W+\overbar{W}$, which implies
\begin{align}
 B^2=W\overbar{W}.
\end{align}
This can be solved numerically to relatively high accuracy, giving
\begin{align}
 A_\mathrm{crit}\simeq 0.0045396037095. \label{A_crit}
\end{align}
This agrees with the results of \cite{Pravda:2000zm}, which were found using a different method.

\section{Numerical solutions for bound orbits} \label{bound}

In this section, we give a few examples of numerical solutions corresponding to geodesics in region (ii) of Fig.~\ref{czv_regions}. These are particles confined to a finite area away from either horizon. So in principle, they can continue to evolve within the region indefinitely up to $\tau\rightarrow\infty$.  We will not attempt an exhaustive classification of numerical solutions here, but rather we show some examples of non-circular bound, co-accelerating orbits. In particular, numerical results suggest that closed, periodic orbits may exist.\par
\begin{figure}
  \begin{center}
   \includegraphics{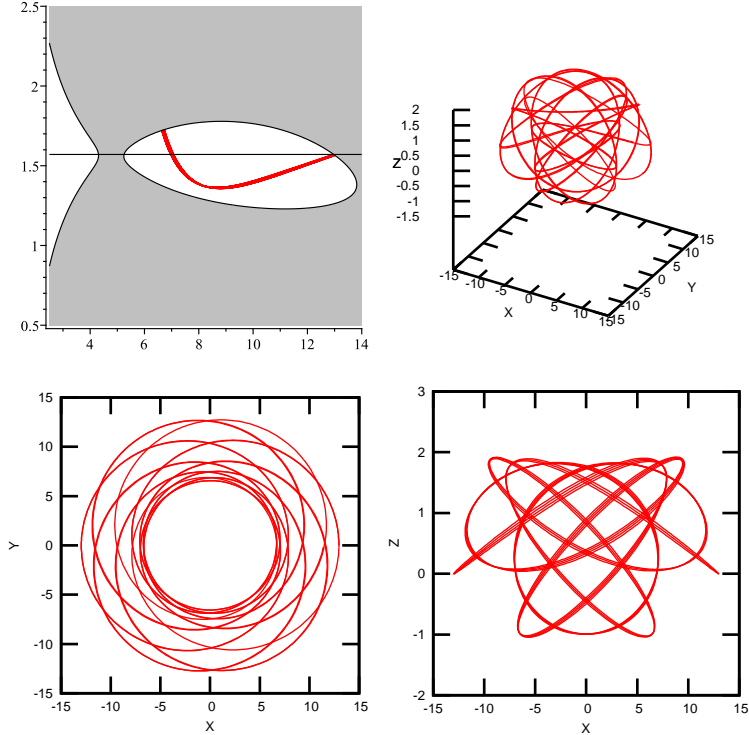}
  \end{center}
  \caption{Geodesics around a black hole with $A=0.001$, $m=1$, for a massive particle of energy $E^2=0.905$, and angular momentum $\Phi^2=13$. The top left figure shows the region in the $r$-$\theta$ plane accessible by the particle. The top right figure shows the geodesics plotted in $(X,Y,Z)$ Cartesian-like coordinates. The bottom left and right are respectively projections to the $(X,Y)$ and $(X,Z)$ plane.}
  \label{plot1}
\end{figure}
\begin{figure}
  \begin{center}
   \includegraphics{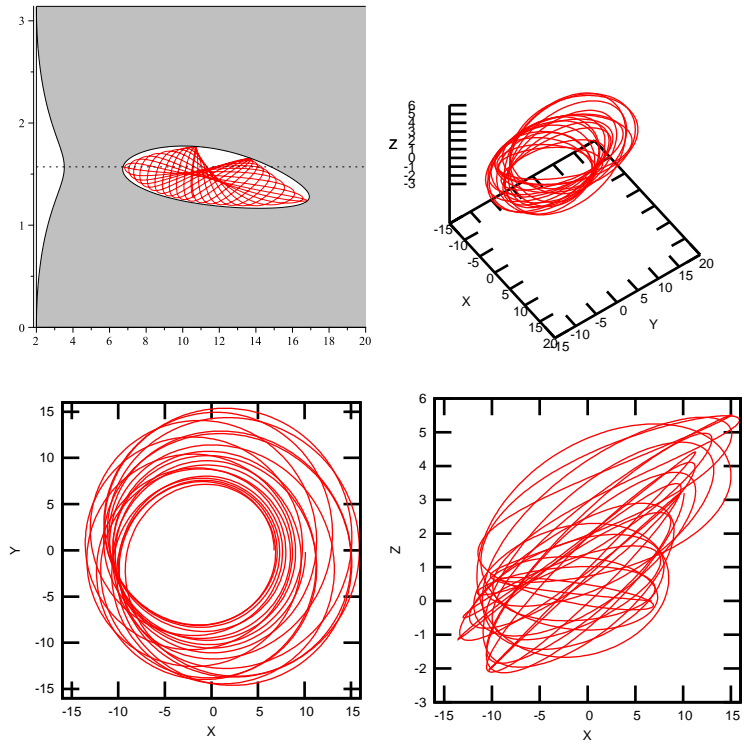}
  \end{center}
  \caption{Geodesics around a black hole with $A=0.001$, $m=1$, for a massive particle of energy $E^2=0.92$, and angular momentum $\Phi^2=14$. The starting point is at $\theta=\pi/2$. The top left figure shows the region in the $r$-$\theta$ plane accessible by the particle. The top right figure shows the geodesics plotted in $(X,Y,Z)$ Cartesian-like coordinates. The bottom left and right figures are respectively projections to the $(X,Y)$ and $(X,Z)$ plane.}
  \label{plot2}
\end{figure}

One such orbit is demonstrated in Fig.~\ref{plot1}, where the geodesics show a mushroom-like profile. It would be interesting to consider the possibility of classifying such orbits in a `periodic table' similar to the periodic tables of Schwarzschild and Kerr geodesics \cite{Levin:2008mq,Grossman:2011ps}.
In other cases, orbits generally do not close exactly. Their trajectories evolve and possibly fill up the region in the $r$-$\theta$ plane defined bounded by $E^2-V_\mathrm{eff}^2$. Fig.~\ref{plot2} shows a case with a non-closed orbit of $A=0.001$, $E^2=0.92$ and $\Phi^2=14$.

\section{Conclusion} \label{conclusion}

In this paper, we have obtained the geodesic equations for the Ricci-flat C-metric and analytical solutions for some special cases have been considered. In the case of zero angular momentum, we have radial geodesics along $\theta=0$ and $\theta=\pi$. In the case of $\theta=0$, it is possible for a timelike particle to remain at unstable equilibrium at a fixed distance away from the black hole. The same was not possible for $\theta=\pi$ as there are no local extrema for the effective potential at the south pole.
\par
The geodesic equations were also used to calculate the stability of Schwarzschild circular orbits under small accelerations along the orbital plane. We have found that, for timelike particles, circular Schwarzschild orbits with $r_0<6m$ are unstable to small accelerations on the system, causing the particle to fall into the black hole. The orbits with $r_0>6m$ are stable as with a small, nonzero $A$, and the particle makes small oscillations about $r_0$, as the perturbation equations takes the form of a harmonic oscillator of frequency $\omega$ with a constant force plus a periodic driving force of frequency $\Omega$. We see that the `natural frequency' $\omega$ is always smaller than $\Omega$, and equality is only reached in the limit $r_0\rightarrow\infty$; therefore we will not see resonance behavior for this motion.
\par
There are analytical solutions to the full geodesic equations ($A$ need not be small) representing circular orbits with its axis parallel to the direction of acceleration. Perturbing around the solutions shows that circular orbits are stable for $mA\leq 0.00454\ldots$, consistent with the numerical results of \cite{Pravda:2000zm} which were found by calculating the local minima of $V_\mathrm{eff}$.
\par
Co-accelerating orbits that are non-circular were found using numerical integration. There are indications showing the existence of closed periodic orbits with a mushroom-like profile. Analogous periodic orbits in Kerr and Schwarzschild spacetimes were found in Ref.~\cite{Levin:2008mq} by calculating the energies and angular momenta that give rational ratios of its orbital frequencies. This approach does not seem possible here as we were not able to separate the Hamilton--Jacobi equation for timelike particles in the C-metric.\footnote{On the other hand, the Hamilton--Jacobi equation for \emph{null} geodesics is easily separable. However the solutions do not yield interesting bound orbits as it is already seen in the effective potential considered in Sec.~\ref{potential}---they either fall into the black hole or fall beyond the acceleration horizon.} Nevertheless, the Hamilton--Jacobi equations for Schwarzschild geodesics are indeed separable, and this is the $A=0$ case of the C-metric equations. It might 
then be possible to find orbital frequencies for small $A$ by methods of canonical perturbation \cite{goldstein,calkin1996}. This possibility will be considered in a future work.

\section*{Acknowledgements}
I would like to thank Edward Teo for useful comments and discussions.

%\bibliographystyle{JHEP}
%\bibliographystyle{ams}
%\bibliography{cmetric}

\end{document}